\input epsf
\input harvmac.tex

\def\capt#1{\narrower{
\baselineskip=14pt plus 1pt minus 1pt #1}}

\lref\hij{Hikihara, T. and Furusaki, A.: Correlation amplitude for
$S={1\over 2}$ XXZ spin chain in the critical region:
Numerical renormalization-group study of an open chain.
Phys. Rev. {\bf B58}, R583-586  (1998)}

\lref\SL{Lukyanov, S.: Low energy effective Hamiltonian
for  the XXZ spin chain. Nucl. Phys. {\bf B522}, 533-549   (1998) }

\lref\sll{Lukyanov, S.: Unpublished}

\lref\lsu{Lieb, E., Schultz, T. and Mattis, D.:
Two soluble models of an antiferromagnetic chain.
Ann. Phys. (NY), {\bf 16}, 407-466 (1961)}

\lref\barry{McCoy, B.M.: Spin correlation functions of the
X-Y model. Phys. Rev. {\bf 173}, 531-541 (1968)}
 
\lref\Baxter{Baxter, R.J.: Exactly solved models
in Statistical Mechanics. London: Academic Press, 1982}

\lref\FLZZZ{Fateev, V., Lukyanov, S., Zamolodchikov, A. and
Zamolodchikov, Al.: Expectation values of local fields
in Bullough-Dodd model and integrable
perturbed conformal field theories.
Nucl. Phys. {\bf B516}, 652-674 (1998)}

\lref\lusher{Luther, A. and Peschel, I.: 
Calculation of critical exponents
in two dimensions from quantum field theory in one dimensions.
Phys. Rev. {\bf B12}, 3908-3917 (1975)}

\lref\Korr{Essler, F.H.L., Frahm, H., Izergin, A.G. and
Korepin V.E.: Determinant representation for correlation functions of
spin-${1\over 2}$ XXX and XXZ Heisenberg magnets. Comm. Math. Phys.
{\bf 174}, 191-214 (1995)}

\lref\MiW{Jimbo, M. and Miwa, T.: Quantum KZ equation
with\ $|q|=1$\ and correlation functions
of  the XXZ model in the gapless regime.
J. Phys. {\bf A29}, 2923-2958 (1996)}

\lref\LZ{Lukyanov, S. and Zamolodchikov, A.:
Exact expectation values of local fields
in quantum sine-Gordon model.
Nucl. Phys. {\bf B493}, 571-587 (1997)}

\lref\affleck{Affleck, I.: Exact correlation amplitude for
the\ $S={1\over 2}$\ Heisenberg antiferromagnetic chain. 
J. Phys. {bf A31}, 4573 (1998)}

\lref\afa{Affleck, I.: Critical behavior 
of two-dimensional systems with
continuous symmetries. Phys. Rev. Lett. {\bf 55}, 1355-1358 (1985)}

\Title{\vbox{\baselineskip12pt\hbox{RU-98-42}
                \hbox{cond-mat/9809254}}}
{\vbox{\centerline{}
\centerline{Correlation amplitude for the XXZ spin chain}
\centerline{in the disordered regime}}}
\centerline{}
\centerline{ Sergei Lukyanov}
\centerline{}
\centerline{Department of Physics and Astronomy,
Rutgers University}
\centerline{ Piscataway,
NJ 08855-0849, USA}
\centerline{and}
\centerline{L.D. Landau Institute for Theoretical Physics}
\centerline{Kosygina 2, Moscow, Russia}
\centerline{}
\centerline{}

\centerline{\bf{Abstract}}

\centerline{}

We proposed an analytical expression for  the amplitude defining
the long distance asymptotic of the correlation function
$\langle\, \sigma^z_k\sigma^z_{k+n}\, \rangle$.

\Date{September, 98}

\eject

One of the most famous model for
1D magnetic is
the XXZ spin chain,
\eqn\xxz{{\bf H}_{XXZ}=-{J\over{2}}
\sum_{k=-\infty}^{\infty}\big(\,
\sigma_{k}^{x}
\sigma_{k+1}^{x}+\sigma_{k}^{y}\sigma_{k+1}^{y}+\Delta\,
(\sigma_{k}^{z}\sigma_{k+1}^{z}-1)\, \big)\ ,}
where $\sigma_{k}^{x}, \sigma_{k}^{y}$ and $\sigma_{k}^{z}$
are the  Pauli matrices
associated with the site $k$.
The energy  spectrum of the model can be studied by 
means of the Bethe ansatz technique (see e.g. Ref.\Baxter\ for a
review).
An exact  calculation of
correlation functions
is a much challenging problem\ \refs{\Korr, \MiW}.
In the disordered regime\foot{
The substitution $J\to -J,\  \Delta\to -\Delta$ transform\
\xxz\ to the  unitary equivalent model. In particular, the chain with
$J>0,\ \Delta=-1$ is unitary equivalent to the $SU(2)$ invariant
antiferromagnetic spin chain.}
\eqn\ystr{ -1\leq\Delta<1\, , \ \ \ \ J>0\ , }
the continuous limit of the chain\ \xxz\ is described by the simple
Conformal Field Theory model (the Gaussian model) and
the qualitative  analysis  of the correlation functions are obtained by
the  Luther-Pershel bosonization 
procedure\ \lusher. The simplest
zero-temperature, equal-time correlators
have the following
leading  behavior\ \lusher,
\eqn\kdjduf{\eqalign{&\langle\, \sigma^x_k\sigma^x_{k+n}\, \rangle=
F\ n^{-\eta}+\ldots\ ,\cr
&\langle\, \sigma^z_k\sigma^z_{k+n}\, \rangle=-{1\over \, \pi^2\, \eta}\ 
n^{-2}+(-1)^n\ A\ n^{-{1\over\eta}}+\ldots\ \ \ {\rm as}
\ \ n\to\infty\ ,}}
where dots stand for subleading terms of the asymptotics. 
The parameter $0<\eta< 1$ in \kdjduf\ is  related with
the anisotropy $\Delta$,
\eqn\hsgdta{\Delta=\cos(\pi\eta)\ .}
The Luther-Pershel approach fails to predict the 
value of the 
correlation amplitudes $F$ and $A$. Recent field-theoretical
results\ \refs{\LZ,\FLZZZ}\ made it possible  to determine
the amplitude $F$\ \LZ,
\eqn\xxzfactor{\eqalign{F=&{1\over 2\, (1-\eta)^2}\
\biggl[\, { \Gamma\big({\eta\over 2-2\eta}\big)\over
2\sqrt\pi\  \Gamma\big({1\over 2-2\eta}\big)}\,
\biggr]^{\eta}\times\cr &
{\rm exp}\biggl\{-
\int_{0}^{\infty} {d t\over t}\ \Big(\,
{ {\rm sinh}\big(\eta t  \big)\over
{\rm sinh}( t)\,
{\rm cosh}\big((1-\eta) t\big )}-
\eta\,   e^{-2 t}\, \Big)\biggr\}\ . }}
This expression was confirmed  numerically in Ref.\hij.

Up to now, only few analytical results  
were known  about the
second amplitude $A$, namely, its values   at the 
``free fermion'' point, $\Delta=0$\ \refs{\lsu, \barry},
\eqn\ksjduyt{A\big|_{\Delta=0}={2\over \pi^2}=0.2026...\ ,}
and at  $\Delta=-{1\over\sqrt2}$\ \sll,
\eqn\kyt{A\big|_{\Delta=-{1\over\sqrt2} }= \Big({2\over 3}\Big)^{
{5\over 3}}\ {4\over \Gamma^4\big({2\over 3}\big)}=0.6053...\ .}

In this letter we propose the exact  formula for the amplitude
$A$,
\eqn\xzior{\eqalign{A=&{8\over \pi^2}\
\biggl[\, { \Gamma\big({\eta\over 2-2\eta}\big)\over
2\sqrt\pi\  \Gamma\big({1\over 2-2\eta}\big)}\,
\biggr]^{{1\over \eta}}\times\cr &
{\rm exp}\biggl\{
\int_{0}^{\infty} {d t\over t}\ \Big(\,
{ {\rm sinh}\big((2 \eta-1) t  \big)\over
{\rm sinh}( \eta t)\,
{\rm cosh}\big((1-\eta) t\big )}-
{2\eta-1\over \eta}\,   e^{-2 t}\, \Big)\biggr\}\ . }}
This function satisfies Eqs.\ksjduyt,\kyt.
One can also check the behavior of $A$
in the limit $\Delta\to-1$,
\eqn\dfjh{A\big|_{\Delta\to-1}\to {2\over \pi} \ \Big({
2\over 1+\Delta}\Big)^{{1\over 4}}\ .}
The divergence is due
to the irrelevant  operator with the scale dimension $2\, \eta^{-1}$
occurring in
the low-energy effective
Hamiltonian of the spin chain\ \afa. 
For the same reason  Eq.\kdjduf\ defines the
leading asymptotics of the correlators only
for
$$ \log(n)\gg {1\over 2-2\, \eta}\ . $$
If $\Delta=-1$, the domain of
validity of\ \kdjduf\ disappears completely.
In order to examine the asymptotic in the vicinity $\Delta=-1$,
we should   perform the standard renormalization
group resummation (see e.g. \refs{\affleck, \SL}).
Using\ \dfjh,\ one can obtain,
\eqn\osuy{\langle\, \sigma^z_k\sigma^z_{k+n}\,
\rangle\big|_{\Delta\to-1}=(-1)^n\ 
\sqrt{{2 \over \pi^3}}\
{2\ \sqrt{-g _{ \perp}}\over n\, (g_{\parallel}- g_{\perp})}\ 
\big(1+O(g) \big)\ ,  }
with
\eqn\hsydt{g_{\parallel}=2\, (1-\eta) \ {1+q\over 1-q}\ ,\ \ \ \ \
g_{\perp}=-4\, (1-\eta) \ {q^{1\over 2}\over 1-q}\ .}
Here $q=q(n,\eta)$ is the solution of the 
equation
\eqn\hsgdr{q\  (1-q)^{{2\over \eta}-2}=
\biggr[\,   {e^{-\gamma-1}\, \eta\,   \Gamma\big({\eta\over
2-2\eta}\big)\over 2\,\sqrt\pi\, n\,   \Gamma
\big({1\over
2-2\eta}\big)}\, \biggl]^{{4\over\eta}-4}  }
and $\gamma=0.5772...$ is the Euler constant.
Now we  take the
limit $g_{\perp}\to 
-g_{\parallel}$, corresponding to
$\Delta\to-1$. The final result 
reproduce the  prediction from\ \refs{\affleck, \SL}\ for
$SU(2)$-invariant
antiferromagnetic spin chain. 
It supports the limiting
behavior\ \dfjh.

The amplitude \xzior\ was also checked against  available
numerical data. In Table and Figure
the numerics from
Ref.\hij\ are compared against\ \xzior.
Notice that the fitting procedure\ \hij\ of the numerical data was
based on  formulas similar to\ \kdjduf
\foot{The authors adapted\ \kdjduf\ for the spin chain with open
boundaries.}. As mentioned above,
the asymptotics\ \kdjduf\ are applicable for very large $n$ only
provided  $\Delta\simeq -1$. 
For $\Delta=-0.9$ the term $\propto n^{2-3/\eta}$,
omitted in\ \kdjduf,\ gives rise a $18\%$ correction
to the value of the correlator
$\langle\, \sigma^z_k\sigma^z_{k+n}\, \rangle$ at $n=100$.
At the same time  the total length of the  chain in\ \hij\ was 200 sites.  
Therefore, the discrepancy between\ \xzior\ and the
numerical data in the vicinity $\Delta= -1$ does not
seem to contradict  our conjecture seriously.

\midinsert

\centerline{\epsfbox{  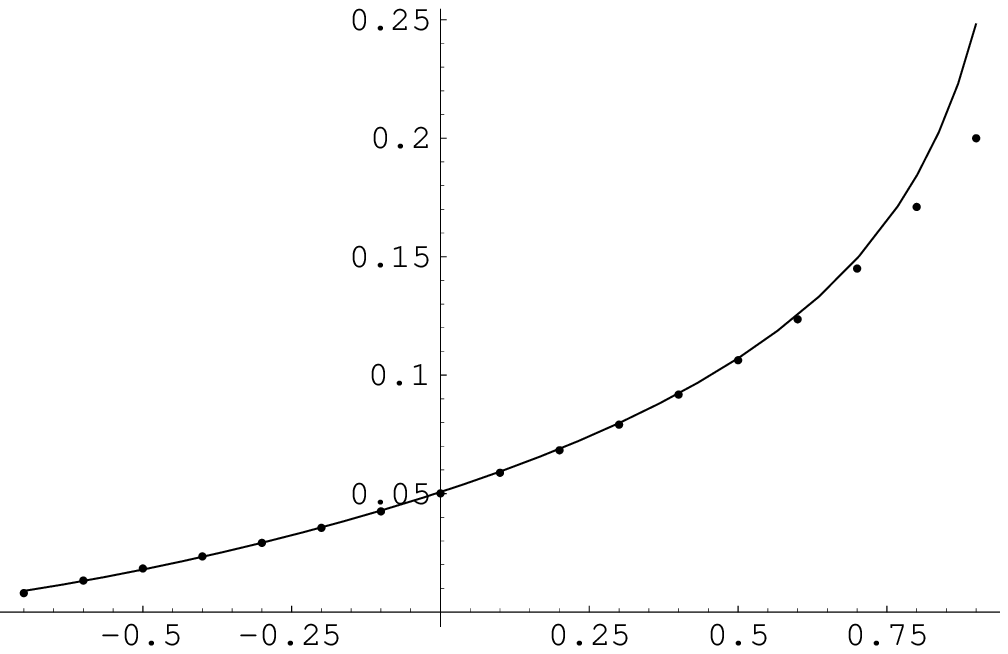 }}
\vskip 5 truemm
\capt{
Figure. The correlation amplitude
$A/4$ from\ \xzior\ (vertical axis)  as a function
of the anisotropy parameter $-\Delta$ (horizontal axis).
The bullets
(see Table) were obtained in\ \hij}.
\vskip 3 truemm
\endinsert
\midinsert
\centerline{
\noindent\vbox{\offinterlineskip
\def\tablerule{\noalign{\hrule}}
\halign{
\strut#&\vrule#\tabskip=1em plus2em&
   #&\vrule#&
   #&\vrule#&
   #&\vrule#
\tabskip=0pt
\cr\tablerule
&& $-\Delta$  && \ $A_{num}$/4   && $A/4$ &
\cr\tablerule
&& -0.7  && 0.008(1)  && 0.00893 &
\cr\tablerule
&& -0.6  && 0.0133(1) && 0.01314  &
\cr\tablerule
&& -0.5  && 0.0184(4) && 0.01795 &
\cr\tablerule
&& -0.4  && 0.0235(2) && 0.02332 &
\cr\tablerule
&& -0.3  && 0.02921(3)&& 0.02924 &
\cr\tablerule
&& -0.2  && 0.03556(3)&& 0.03574 &
\cr\tablerule
&& -0.1  && 0.0425(2) && 0.04285 &
\cr\tablerule
&& -0.0  && 0.0501(5) && 0.05066 &
\cr\tablerule
&&  0.1  && 0.0588(3) && 0.05929 &
\cr\tablerule
&&  0.2  && 0.0683(6) && 0.06891 &
\cr\tablerule
&&  0.3  && 0.0791(8) && 0.07978 &
\cr\tablerule
&&  0.4  && 0.0918(9) && 0.09231 &
\cr\tablerule
&&  0.5  && 0.1063(9) && 0.10713 &
\cr\tablerule
&&  0.6  && 0.1236(5) && 0.12539 &
\cr\tablerule
&&  0.7  && 0.145(1)  && 0.14930 &
\cr\tablerule
&&  0.8  && 0.171(5)  && 0.18414 &
\cr\tablerule
&&  0.9  && 0.20(1)   && 0.24844 &
\cr\tablerule
\noalign{\smallskip} }
}
}
\bigskip
\capt{Table. The correlation amplitude $A_{num}/4$ was estimated in the
paper\ \hij. $A/4$ follows from Eq.\xzior }
\endinsert
\centerline{}

\centerline{\bf Acknowledgments}

\vskip0.3cm

I am grateful A.B. Zamolodchikov for interesting discussions.
The research  is supported in part by the DOE grant \#DE-FG05-90ER40559.

\listrefs

\end